\documentstyle[prb,aps,epsf,epsfig,twocolumn]{revtex}
\def\gtappr{{{\lower4pt\hbox{$>$} } \atop \widetilde{ \ \ \ }}}
\def\ltappr{{{\lower4pt\hbox{$<$} } \atop \widetilde{ \ \ \ }}}
\def\be{\begin{equation}}
\def\ee{\end{equation}}
\def\bea{\begin{eqnarray}}
\def\eqa{\begin{eqnarray}}
\def\eea{\end{eqnarray}}

\def\dg{{^{\dag}}}

\newlength{\wdy}
\newlength{\hty}
\newlength{\widf}
\begin{document}
\draft
\twocolumn[\hsize\textwidth\columnwidth\hsize\csname@twocolumnfalse%
\endcsname
\title{Quantum Reciprocity Conjecture for the Non-Equilibrium Steady State}
\author{P. Coleman and W. Mao}
\address{Center for Materials Theory, Rutgers University, Piscataway, NJ 08854, USA
}

\date{\today}
\maketitle
\vskip 0.5 cm
\begin{abstract}
A consideration of the lack of history dependence in the non-equilibrium
steady state of a quantum system leads us to conjecture that in such
a system, there is a set of quantum
mechanical observables whose retarded response 
functions are insensitive to
the arrow of time, and which consequently satisfy a quantum analog of
the Onsager reciprocity relations. 
Systems which satisfy this conjecture can be described by an effective
Free energy functional. We demonstrate that the conjecture holds in 
a resonant level model of  a multi-lead quantum dot. 
\end{abstract}

\vspace{1 cm}

\vskip -0.4 truein
\vspace{0.5 cm}
]
\narrowtext
%
%
%
%
%
%
Although the fundamental principles of thermal equilibrium 
were established by Boltzmann more than a century ago, their
generalization to the non-equilibrium steady state has proved elusive.
The non-equilibrium steady state is
thought to be defined by a set of  characteristic 
variables such as the  current, the thermal and chemical potential
gradient and as such, it is expected 
to be independent  of the history of how it was prepared. 
This has led to the notion that 
general principles should govern the instantaneous properties of the
steady state. 
One recurring idea is that a generalized free energy functional might apply to the
non-equilibrium steady 
state\cite{rayleigh,Onsager:1931a,Onsager:1931b,hershfield,Christen,derrida}.  This was first speculated by 
Rayleigh in the late 19th century.\cite{rayleigh}
Onsager\cite{Onsager:1931a,Onsager:1931b} later used his
reciprocity relations to support this conjecture, but the idea has
remained controversial to the present day.

Non-equilibrium steady state behavior plays an important
role in electronic transport theory, and becomes particularly important
in driven nano devices, such as a d.c. biased quantum
dot. \cite{Intro} Variants on Rayleigh's approach 
would be invaluable in this new context,  and might provide
an important 
first step along the road to Boltzmann's approach 
the non-equilibrium steady state.\cite{hershfield,Christen,derrida}

Recent work on non-equilibrium hydrodynamics
has shown how Onsager's reciprocity relations can be generalized
to the non-equilibrium steady state. \cite{McLennan,Dufty} 
This motivates us to re-examine 
Onsager's reciprocity relations in the context of non-equilibrium
quantum physics. 
By considering the history independence 
of the non-equilibrium steady state, we are led to conjecture
that Onsager's reciprocity theorem continues within a limited
class of quantum variables, in the non-equilibrium
steady state.  Within this restricted class of variables, 
the concept of a Free energy can be used to
describe the steady state of non-equilibrium quantum systems. 

\begin{figure}      
\vspace{-0.5truecm}  
\center      
\centerline{\epsfxsize=2.0in      
\epsfbox{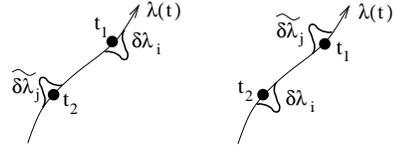}}  
\vspace{-0.1truecm}      
\begin{minipage}[t]{8.1cm}      
\caption{Two variations in the path P
where the increments in  $\lambda _{j}$ at times $t_{2}$
and $t_{1}>t_{2}$ are interchanged. 
}    
\label{vary}     
\end{minipage}      
\end{figure}

The lack of history dependence of the 
equilibrium steady state
means that 
the 
work done on the system by coupling 
various internal degrees of freedom 
$\hat  A_{i}$ ($i=1,n$) 
to corresponding external ``forces''  $\lambda_{i} (t)$, 
\[
 W = 
\sum _{i}\int_{P} 
\langle A_{i}
(t)\rangle d\lambda_{i} (t) ,
\]
does not depend on the path $P$ over which the 
$(\lambda_{i})$ are adiabatically incremented 
to their final value.  If we increment $\lambda _{j} (t)$ at two 
different times $t_{2}$ and $t_{1}>t_{2}$, we may do it two ways, 
illustrated in Fig.  (\ref{vary}).
\noindent 
In the first  variation
$\lambda _{i} (t_{1})\rightarrow
\lambda _{i} (t_{1})+\delta \lambda _{i}$ 
and 
$\lambda _{j} (t_{2})\rightarrow
\lambda _{j} (t_{2})+\widetilde{\delta \lambda  } _{j}$, whereas in
the second the variations are reversed $\delta \lambda
_{j}\leftrightarrow \delta \tilde{\lambda } _{j}$.
The second-order change in the
work done along both paths must be equal, i.e 
\begin{equation}\label{}
{\delta ^{2} W }= 
\delta \lambda _{i}\widetilde{\delta \lambda _{j}}
\left(\frac{\delta \langle A_{i} (t_{1}) \rangle }{\delta \lambda _{j} (t_{2})} \right)
= 
\delta \lambda _{i}\widetilde{\delta \lambda _{j}}
\left(\frac{\delta \langle A_{j} (t_{1}) \rangle }{\delta \lambda _{i} (t_{2})} \right)
\end{equation}
from which if follows that 
\begin{eqnarray}\label{}
\frac{\delta \langle A_{j}(t_{1})\rangle }{\delta \lambda _{i} (t_{2})} -
\frac{\delta \langle A_{i} (t_{1})\rangle }{\delta \lambda _{j} (t_{2})} &=& 0.
\end{eqnarray}
We can relate these functional derivatives
to the corresponding response functions, 
\begin{eqnarray}\label{}
\frac{\delta \langle A_{j} (t)\rangle }{\delta \lambda _{i} (t')} &=&
-i\langle [A_{j} (t),A_{i} (t')]\rangle \theta (t-t')
\end{eqnarray}
from which it follows that
\begin{eqnarray}\label{}
-i\langle [A_{j} (1),A_{i} (2)]\rangle \theta (1-2)&=&
-i\langle [A_{i } (1), A_{j} (2)]\rangle \theta (1-2).
\end{eqnarray}
These are the quantum generalization of
Onsager's reciprocity relations\cite{Onsager:1931a,Onsager:1931b}.
The relations are understood to hold only in the long-time limit
corresponding to a slow adiabatic variation of the
source
terms. 
Onsager identified relations  
with the
microscopic reversibility of the equations of motion  and the absence
of any ``arrow of time'' in thermal equilibrium.  This derivation shows
how  reciprocity is directly related  to a lack of history
dependence. Since our proof makes no reference to thermal equilibrium, 
it offers the intriguing prospect
of an extension to the non-equilibrium steady state. 

To extend the discussion away from thermal equilibrium, we consider a tiny system ``$S$'',  which may be a
quantum dot\cite{Intro,Glazman:2000,us}, a quantum wire\cite{devoret}, or other small system that is 
coupled to two
very large baths of electrons (``leads'') at different chemical potentials
$\mu_{L}$
and $\mu_{R}$ where $\mu_{L}> \mu_{R}$.   The entire coupled system
is completely isolated from the outside world. 
\begin{figure}      
\vspace{-0.5truecm}  
\center      
\centerline{\epsfxsize=2.0in      
\epsfbox{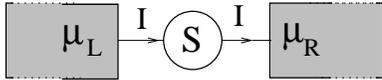}}  
\vspace{-0.1truecm}      
\begin{minipage}[t]{8.1cm}      
\caption{The non-equilibrium steady state is obtained by adiabatically
connecting system S to two heat baths at chemical potentials $\mu
_{L,R}$. 
}    
\label{system}     
\end{minipage}      
\end{figure} 
\noindent 
If we connect $S$ to the leads
at time $t=0$, then after an equilibration time
$\tau _{1}$,  the system will arrive at a steady state where a current 
flows from the left to the right-hand lead. 
This state
persists for a long time $\tau _{2} (L)$
until a substantial fraction of the additional electrons on the left
lead have flowed into the right lead.  The time $\tau _{2} (L)$
will diverge rapidly as $L\rightarrow \infty $, which permits us to define
the steady state value of some variable $\hat
A$ as
\[
\langle A\rangle  = \lim_{L\rightarrow \infty }\langle A (t)\rangle 
\]
with the understanding that $\tau _{2} (L)>>t>>\tau _{1}$. 

Suppose
the steady state is arrived by adiabatically turning on an interaction
$H_{I}= gh_{I}$ between the leads, and by coupling source terms
$\lambda_{j}$ to various quantities $A_{j}$ which are 
localized within $S$.
Since the combined system is closed, when  we adiabatically change these variables 
the amount of work done
in reaching the steady state is simply 
the change in the total energy of the system
\[
W_{NE} = \int \langle h_{I} (t)\rangle  dg (t) +  \langle A_{i}
(t)\rangle d\lambda_{i} .
\]
If the work done $W_{NE}$ is  independent
of the path by which $g$ and the $\lambda _{j}$ reach 
their final values, then we can use the previous proof to show that
the corresponding variables satisfy a quantum reciprocity relation. 
The converse will also hold true. 
This motivates the ``Quantum Reciprocity
Conjecture'':
\begin{quote}
{\sl In the non equilibrium steady state, the set of quantum
mechanical observables
contains a non-trivial subset ${\mathcal{P}}$  of ``protected''
quantum observables ${\mathcal{P}}=\{a_{1},a_{2}\dots ,a_{n} \}$ whose
correlation functions in the steady state
are insensitive to the arrow of time, and which consequently satisfy a 
quantum mechanical analog of the Onsager reciprocity relations
\[
\langle [a (1),b (2)]\rangle = \langle [b (1),a (2)]\rangle , \qquad (a,b\
\in {\mathcal{P}}).
\]}
\end{quote}
Of course we do not expect
the reciprocity relation to extend to {\sl all} variables,
as it does
in thermal equilibrium, because this would mean that the arrow of
time is completely invisible. 

Consider the retarded and advanced
Green functions between protected variables, 
\begin{eqnarray}\label{}
G^{( R,A)}_{ab} = \mp i \langle  [a (1),b (2) ]\rangle \theta_{\pm} (t_{2}-t_{1})
\end{eqnarray}
where $\theta _{\pm} (t)= \theta (\pm t)$. Since $a$ and $b$ are
hermitian, these are real
functions ($G^{R,A} (t)=[G^{R,A} (t)]^{*}$). The 
conjectured Onsager relations mean that in the steady state, 
they also satisfy
\begin{eqnarray}\label{}
G^{ R}_{ab} (t_{2}-t_{1})&=& G^{ A}_{ab}(t_{1}-t_{2}),\cr
G^{( R,A)}_{ab} (t_{2}-t_{1})&=&G^{( R,A)} _{ba}(t_{2}-t_{1}),
\end{eqnarray}
where the order of the subscripts and time variables is important.
If we write $G^{R}(t_1-t_2)= \left[G^{R}(t_1-t_2)\right]^*$ in the first relation, and then
Fourier transform, 
we obtain the more familiar result
\[
G^{A}_{ab} (\omega )=G^{R}_{ab} (\omega)^{*}
\]
which means that the 
retarded and advanced Green functions of protected variables  share
the same spectral decomposition
\[
G^{{(R,A)}}_{ab} (\omega ) = \int \frac{dE}{\pi} \frac{1}{\omega
-E \pm i \delta }A_{ab} (E)
\]
where $A_{ab} (E)= \pm{\rm Im }[G^{(A,R)}_{ab} (E)]$.

\begin{figure}      
\vspace{-0.5truecm}  
\center      
\centerline{\epsfysize=1.2in      
\epsfbox{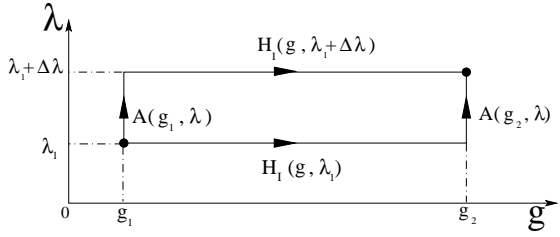}}  
\vspace{-0.1truecm}      
\begin{minipage}[t]{8.1cm}      
\caption{Two paths for turning on the interaction and source
terms. 
}    
\label{Path}     
\end{minipage}      
\end{figure}

Provided that the set of protected quantum variables includes the interaction
$H_{I}=gh_{I}$, 
then we can define an effective Free energy
from the virtual work done $W_{NE}$ in reaching the steady state.
Suppose we evaluate $W_{NE}$ along the two paths shown in Fig. \ref{Path}. 
Since $W_{NE}$ is the same along both paths, 
for small $\Delta\lambda$ we have
\eqa
&&A(g_1,\lambda)\Delta\lambda+\int_{g_1}^{g_2}\frac{dg'}{g'}H_I(g',\lambda+\Delta\lambda)dg'\cr
=&&
A(g_2,\lambda)\Delta\lambda+\int_{g_1}^{g_2}\frac{dg'}{g'}H_I(g',\lambda)dg',
\eea
so that 
\eqa
\Delta
A=A(g_2,\lambda)-A(g_1,\lambda)=\frac{\partial}{\partial\lambda}
\Delta F\label{Reciprocity}
\eea
where 
\begin{equation}\label{}
\Delta F=\int_{g_1}^{g_2}\frac{dg'}{g'}H_I(g',\lambda).
\end{equation}
Thus  if reciprocity holds, the change in the 
variables $\{A_{j} \}$
associated with a change in the coupling constant $g$ can be evaluated
as derivatives of a {\sl single} Free energy variable $\Delta F$. 

We now illustrate the correctness of this conjecture in 
a simple  non-interacting model. We consider a single 
\noindent resonant 
level in a quantum dot 
carrying a D.C. current between 
two or more leads according, where  the Hamiltonian
$H=H_{0}+H_{I}$ and 
\bea
H_0&&=\sum_{\alpha,{\bf k}\sigma } \epsilon({\bf k}) c^{\dagger}_{\alpha,{\bf
k}\sigma }c_{\alpha,{\bf k}\sigma }+\sum _{\sigma }\epsilon_{d\sigma
}d^{\dagger}_{\sigma }d_{\sigma },\cr
H_I&&=J\sum_{\alpha,{\bf k}}\left[\gamma
_{\alpha}c^{\dagger}_{\alpha,{\bf k}\sigma }d_{\sigma }
 +H.C.\right].
\nonumber
\eea
Here $\alpha =1, N$ labels the leads, each one characterized by
a distinct chemical potential $\mu _{\alpha }$, $\epsilon _{d\sigma
}=\epsilon _{d}-\sigma B$ is the energy of the localized state in the dot in a
magnetic field $B$, 
$J$ is the overall coupling constant and $\gamma _{\alpha }$
is a parameter which sets the relative 
strength of hybridization with the $\alpha $ lead. 
This is an exactly solvable problem, and has well known results\cite{Wingreencurrent} found
by the Keldysh method. 

As a first step, by comparing the retarded and advanced 
correlation functions, we are able to explicity confirm that 
the interaction, together with the dot magnetization
$M$ and occupancy $n_{d}$, form a set of protected variables 
$\{H_{I},M,n_{d} \}$ which satisfy reciprocity and for which a Free energy 
functional can be defined.  

For example, to confirm the relation
\eqa
\langle[H_I(t_1),n(t_2)]\rangle=\langle[n(t_1),H_I(t_2)]\rangle,
\eea
we compare the retarded and advanced Green functions:
\eqa
G^R_{H_In}(\omega)={\rm Tr}\sum_{\alpha}J\gamma _{\alpha}\int&&\frac{d\epsilon}{2\pi}\left[{\mathcal{G}}_{dd^{\dagger}}(\epsilon)(i\tau_1){\mathcal{G}}_{c_{\alpha}d^{\dagger}}(\epsilon+\omega)\right.\cr
&&\left.+{\mathcal{G}}_{dc^{\dagger}_{\alpha}}(\epsilon)(i\tau_1){\mathcal{G}}_{dd^{\dagger}}(\epsilon+\omega)\right]
\eea
and
\eqa
G^A_{H_In}(\omega)={\rm Tr}\sum_{\alpha}J\gamma _{\alpha}\int&&\frac{d\epsilon}{2\pi}\left[{\mathcal{G}}_{c_{\alpha}d^{\dagger}}(\epsilon+\omega)(i\tau_1){\mathcal{G}}_{dd^{\dagger}}(\epsilon)\right.\cr
&&\left.+{\mathcal{G}}_{dd^{\dagger}}(\epsilon
+\omega)(i\tau_1){\mathcal{G}}_{dc^{\dagger}_{\alpha}}(\epsilon)\right].\eea
where the $G_{ab}$ refer to the Larkin-Ovchinikov matrix Greens
function\cite{LO,Rammer} between electron fields $a$ and $b$ and the trace is over
Keldysh indices. 
By writing these expressions out explicity, we are able to explicitly confirm
that they are related by complex conjugation, 
$G^R_{H_In}(\omega)=[G^A_{H_In}(\omega)]^*$,
from which reciprocity between $n_d$ and $H_I$ holds.
A similar method enables us to check that 
\eqa
\langle \left[H_I(t_1),M(t_2)\right]\rangle=\langle \left[M(t_1), H_I(t_2)\right]\rangle.
\eea
The correlation function between $M$ and $n_{d}$ identically vanishes,
trivially satisfying reciprocity.   

We now confirm that an effective Free energy correctly determines the
occupancies and magnetization. The expectation value of the
interaction energy determined by the equal time Keldysh Green
functions
between the conduction and dot electron, given by 
\[
\langle H_{I}\rangle = J \sum_{\alpha ,\sigma  }\gamma _{\alpha }\int
\frac{d\omega }{4\pi i}\left[ G^{K}_{d_{\sigma } c\dg _{\alpha }} (\omega )+
G^{K}_{c_{\alpha } d\dg _{\sigma  }} (\omega )\right]
\]
After integrating over the coupling constant we obtain
\bea
\Delta F_{\rm eff}=&&\int_0^J\frac{dJ'}{J'}\langle H_I\rangle\cr
=&&\sum_{\alpha ,\sigma}  \frac{2\gamma^{2} _{\alpha } }{\pi}{\rm{Re}}\left[
-2\pi T\log\Gamma\left(\frac{1}{2}+\frac{\epsilon_{d\sigma
}+i\Delta-\mu_\alpha }{2\pi iT}
\right)\right.\nonumber\\
&&+\left.2\pi T\log\Gamma\left(\frac{1}{2}+\frac{\epsilon_{d\sigma
}-\mu_\alpha }{2\pi iT}
\right)+\Delta\ln\left(\frac{D}{2\pi T
}\right)
\right],
\eea
where $\Delta=\sum_{\alpha}\pi\rho ( J\gamma _{\alpha})^{2}$. 
The expectation value of local state occupancy $n_{d}$ and
magnetization $M$ are then
\begin{eqnarray}\label{}
\langle n_d\rangle&=&\frac{\partial \Delta F_{\rm eff}}{\partial
\epsilon_d} +c_{1},\cr
\langle M\rangle&=&-\frac{\partial \Delta F_{\rm eff}}{\partial
B} +c_{2},
\end{eqnarray}
where the constant terms gives the limiting value of the occupancy and
magnetization when $J\rightarrow 0$. We can fix these constants 
by using 
the condition that $\langle n_d\rangle\rightarrow 1$ and $\langle
M\rangle \rightarrow 0$
as $\Delta\rightarrow \infty$. 
which
gives 
\begin{eqnarray}\label{}
\langle n_d\rangle&=&1+\sum_{\alpha,\sigma  }\frac{\gamma^{2} _{\alpha
}}{\pi}{\rm{Im}}\left[\psi\left(\frac{1}{2}+\frac{\epsilon_{d\sigma
}-\mu_\alpha +i\Delta}{2\pi iT}\right)\right],\cr
\langle M\rangle &=&\sum_{\alpha ,\sigma}\frac{\gamma ^{2}_\alpha
}{\pi}\sigma{\rm{Im}}\left[ \psi\left(\frac{1}{2}+
\frac{\epsilon_d+\sigma B-\mu_\alpha +i\Delta}{2i\pi T}\right) \right].
\end{eqnarray}
\begin{figure}      
\vspace{-0.5truecm}  
\center      
\centerline{\epsfysize=1.7in      
\epsfbox{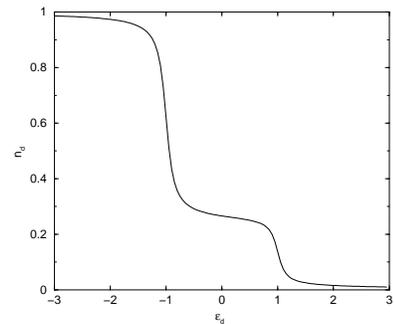}}  
\vspace{-0.1truecm}      
\begin{minipage}[t]{8.1cm}      
\caption{Distribution function of $n_d$ as a function of $\epsilon_d$. $\mu_1=1$, $\mu_2=-1$, $\lambda_1=0.75$, $\lambda_2=0.25$, $\Delta=0.01$, and $T=0.001$.
}    
\label{Dense}     
\end{minipage}      
\end{figure}  
\noindent Both results can  be independently confirmed by direct calculation from the
Keldysh Green functions. It is remarkable that 
the derivative of a single Free energy functional reproduces
the results of two separate Keldysh calculations, even though a
D.C. current is flowing through the dot. 
It is interesting to see that even at the zero coupling limit, 
the occupancy and magnetization of the ``dot'' 
a non-thermalized form, and depends on the ratios between hybridization
$\gamma _{\alpha }$. The non-thermal function $n_{d} (\epsilon _{d})$ is reminiscent
of the occupancy  observed in quantum wire experiments.\cite{devoret} Here the parameters $\lambda_i$ play the similar role of distances between the measured point and leads in the experiment.

It is instructive to examine the the magnetization in the two-lead case 
which for zero temperature is 
\eqa
\chi(B,\Delta)=\frac{2\Delta(B^2+\Delta^2+V^2)}{\pi((B-V)^2+\Delta^2)((B+V)^2+\Delta^2)}
\eea
whilst for $\Delta\rightarrow 0$,
\eqa
\chi(B,T)=\frac{1}{4T}\left[{\rm sech}^2\left(\frac{B+V}{2T}\right)+   {\rm sech}^2\left(\frac{B-V}{2T}\right)\right].
\eea
In both limits, the bias voltage dramatically reduces the 
susceptibility and at a
finite voltage the $T=0$  magnetic susceptibility in the limit
of $J\rightarrow 0$ is always zero.
Non thermal magnetizations of this kind have recently 
obtained in the zero order magnetic susceptibility calculation for quantum dot\cite{KG,PH,Roschnew}. 
Can we extend the set of ``protected'' variables to include
other quantities of interest, such as the current or the spin current?
The answer appears to be ``no''. 
When we directly compare the retarded and advanced
correlators involving any operator that involves the lead electrons,
{\sl other } than $H_{I}$, we find that they are not complex
conjugates. This means that we can not change the ratio of the
couplings $\gamma _{\alpha }$ as we turn on the interaction, for to do this
would be to introduce new variables which do not satisfy
the Onsager reciprocity relation with $h_{I}$. 

The validity of 
our conjecture in more complex systems
is an  open issue. We can not prove that 
reciprocity is stable against the presence of interactions within the
dot, but we have circumstantial support  for this idea.
The above methods can be used in  the large $N$ limit of the
infinite $U$ Anderson model to examine how the
mean-field equations evolve away from equilibrium.  We have also
compared the local susceptibility in 
the non-equilibrium Kondo problem obtained using the reciprocity
conjecture
with that obtained using Majorana techniqes.\cite{tobepublished} 
An interesting
recurring feature of these calculations, is the appearance of non-thermal
distribution functions in the limit that the coupling with  the 
leads is taken to zero. In interacting systems,
these limiting distribution functions will need 
need to be computed
self-consistently from the limiting form of the Dyson equation, before
the change in Free energy can be computed.\cite{cm}

In conclusion, we have examined the idea that the principle of virtual
work can be extended to the non-equilibrium steady state of quantum
systems. This has led us to conjecture the existence of a class of
steady state variables which satisfy the quantum generalization of
Onsager's reciprocity relation out of equilibrium. If this conjecture
holds, then the notion of a free energy can be extended to the quantum
non-equilibrium steady state, permitting the expectation values of
steady state variables to be computed as derivatives of a free energy
functional. This idea works for the simplest possible example, and
leaves open the possibility that it will apply to more complex and
interesting interacting situations.


We  wish to thank  
Chris Hooley, David Langreth, Joel Lebowitz and Olivier  Parcollet,  
for the many lively discussions  at the Rutgers Center for
Materials Theory that led to this paper.  
This work was supported by DOE grant DE-FG02-00ER45790.


\begin{thebibliography}{99}
\bibitem{rayleigh}Lord Rayleigh, Proc. Math. Soc London {\bf 4}, 357,
[363], (1873); Theory of Sound (London, MacMillan Co, 1st Ed 1877),
Vol 1, p 78 (2d ed 1878), Vol 1, p 102. 
\bibitem{Onsager:1931a}L. Onsager, 
Physical Review 37,405(1931)
\bibitem{Onsager:1931b}L. Onsager, 
Physical Review 38, 2265(1931).
\bibitem{hershfield}S. Hershfield, Phys. Rev. Lett {\bf 70}, 2134, (1993).
\bibitem{Christen}T. Christen, Phys. Rev. B {\bf 55}, 7606, (1997).
\bibitem{derrida}B. Derrida, J. L. Lebowitz and E. R. Speer,
Phys. Rev. Lett. 87, 1506001 (2001).
\bibitem{Intro} M. A. Kastner, Rev. Mod. Phys. 64, 849 (1992); R. C. Ashoori, Nature 379, 413 (1996); L. P. Kouwenhoven and C. Marcus, Physics World 11, 35 (June 1998).
\bibitem{McLennan}J.A. McLennan, Phys. Rev. A {\bf 10}, 1272 (1974).
\bibitem{Dufty}J.W. Dufty and J.M. Rub{\'\i}, Phys. Rev. A {\bf 36}, 222 (1987).
\bibitem{Glazman:2000} A. Kaminski, Yu. V. Nazarov, and L. I. Glazman,
Phys. Rev. {\bf B} {\bf 62}, 8154(2000).
\bibitem{us}P. Coleman, C. Hooley and O. Parcollet, Phys. Rev. Lett. {
\bf 86} 4088 (2001). 
\bibitem{devoret}H. Pothier et al., Phys. Rev. Lett. {\bf 79}, 3490
(1997);
\bibitem{Wingreencurrent} N. Wingreen A-P Jauho and Y. Meir, Phys. Rev. {\bf B48}, 8488(19
93).
\bibitem{LO}Larkin, A. I., and Yu.N. Ovchinnikov, 1975, Zh. Eksp. Teor. Fiz. 68, 1915 (Sov. Phys.-JETP 41, 960
(1975)). 
\bibitem{Rammer} For a review of Keldysh method, see J. Rammer and H. Smith, Rev. Mod. Phys., {\bf 58}, 323(1986).
\bibitem{KG}A. Kaminski and L. Glazman, private communication (2001).
\bibitem{PH}O. Parcollet and C. Hooley, to be published {\tt
cond-mat/0202425}.
\bibitem{Roschnew}A. Rosch, J. Passke, J. Kroha and P. W{\"o}lfle,
condmat/0202404 (2002).
\bibitem{tobepublished}W. Mao et al, to be published (2002).
\bibitem{cm}The results of 
P. Coleman and W. Mao, cond-mat/0203001v1 did not take this fact into
account and will be revised in a future posting.
\end{thebibliography}
\end{document}